# The Thomas-Fermi quark model and mesonic matter[a)]

Suman Baral[b)] and Walter Wilcox[c)]

*Department of Physics, Baylor University*

The first results of a new application of the Thomas-Fermi statistical quark model to mesonic matter is presented. Interesting aspects of the theory are discussed, distinguishing such states from baryonic matter. A major motivation of this study is the tetraquark states discovered by the Belle and other collaborations and the possibility that stable multi-quark families of such states exist. Similar to our previous baryonic study, we use a two-inequivalent wave function approach to investigate aspects of multi-quark matter. We think of our model as a tool for quickly assessing the characteristics of new, possibly bound, particle states of higher quark number content, which can not yet be examined by lattice methods.

## I. INTRODUCTION

One of the long standing issues which has intrigued particle physicists is the possible existence of exotic states of baryons of more than three quarks, or mesons of two or more quark/antiquark pairs, or mixed combinations. Thanks to the observations made by Belle[1], BESIII[2], LHCb[3] and other collaborations, many of these multi-quark states have now been observed. Lattice QCD can successfully address this situation for smaller number of quarks. However, when one goes to much higher quark content extensive computation needs to be done, which is extremely intensive and expensive. So, there is a need for models that can extend beyond present day lattice calculations and lead in the right direction. The Thomas-Fermi (TF) statistical model has been amazingly successful in the explanation of atomic spectra and structure as well as in nuclear applications. Our group has adopted the TF model and adapted it for multi-quarks[4]. It has already been used to investigate multi-quark states of baryons.[5] In this paper, we have extended the TF quark model to mesonic states in order to investigate the stability of families built from some existing mesons and observed new exotic states. We will build these states and examine the energy slopes to determine family stability.

## II. MATHEMATICAL PRELIMINARIES

### A. Residual coulombic coupling

The color couplings of quarks and antiquarks in our model originates from the coulombic interactions expected at the classical level[6]. In the following we

---
a)Research partly supported by the Baylor University Research Committee.
b)Electronic mail: suman_baral@baylor.edu
c)Electronic mail: walter_wilcox@baylor.edu

define $\eta$ to be the number of quark/antiquark pairs in the meson, which is assumed to be a color singlet. In addition, $g$ represents the strong coupling constant. The types of interactions between the particles can then be categorized as:

**Color-Color Repulsion (CCR)**
Interactions between same colors which is repulsive with coupling constant $4/3g^2$. The interactions are red-red ($rr$), green-green ($gg$) and blue-blue ($bb$).

**Color-Color Attraction (CCA)**
Coupling constant is $-2/3g^2$. The interactions are $rb$, $rg$, $bg$, $br$, $gr$, and $gb$.

**Color-Anticolor Repulsion (CAR)**
Coupling constant is $2/3g^2$. The interactions are $r\bar{b}$, $r\bar{g}$, $b\bar{g}$, $b\bar{r}$, $g\bar{r}$, and $g\bar{b}$.

**Color-Anticolor Attraction (CAA)**
Coupling constant is $-4/3g^2$. The interactions are $r\bar{r}$, $b\bar{b}$ and $g\bar{g}$.

**Anticolor-Anticolor Repulsion (AAR)**
Coupling constant is $4/3g^2$. The interactions are $\bar{r}\bar{r}$, $\bar{b}\bar{b}$ and $\bar{g}\bar{g}$.

**Anticolor-Anticolor Attraction (AAA)**
Coupling constant is $-2/3g^2$. The interactions are $\bar{r}\bar{b}$, $\bar{r}\bar{g}$, $\bar{b}\bar{g}$, $\bar{b}\bar{r}$, $\bar{g}\bar{r}$, and $\bar{g}\bar{b}$.

Each of the $\eta$ pairs must be a color singlet and each singlet can be achieved three different ways, red-antired, green-antigreen and blue-antiblue. Assuming equal probability of occurrence, any meson with $\eta$ pairs of quarks might be formed $3^\eta$ number of ways. We carefully counted all the possible interactions out of $3^\eta$ color-anticolor pairs and calculated the probability. Interaction types along with their respective probabilities are listed in the Table I.

| Interaction type | Symbol | Coupling | Interaction probability |
|---|---|---|---|
| CCR | $P_{ii}$ | $\frac{4}{3}g^2$ | $\frac{(\eta-1)}{18(2\eta-1)}$ |
| CCA | $P_{ij}, i \neq j$ | $-\frac{2}{3}g^2$ | $\frac{\eta-1}{9(2\eta-1)}$ |
| CAR | $\bar{P}_{ij}, i \neq j$ | $\frac{2}{3}g^2$ | $\frac{2(\eta-1)}{9(2\eta-1)}$ |
| CAA | $\bar{P}_{ii}$ | $-\frac{4}{3}g^2$ | $\frac{(\eta+2)}{9(2\eta-1)}$ |
| AAR | $\bar{\bar{P}}_{ii}$ | $\frac{4}{3}g^2$ | $\frac{(\eta-1)}{18(2\eta-1)}$ |
| AAA | $\bar{\bar{P}}_{ij}, i \neq j$ | $-\frac{2}{3}g^2$ | $\frac{(\eta-1)}{9(2\eta-1)}$ |

TABLE I. The coupling constants and probabilities for certain types of quark and antiquark interactions in mesons.

From Table I we can see that $\sum_{i \leq j}^{3} P_{ij} + \bar{P}_{ij} + \bar{\bar{P}}_{ij} = 1$. Also, if we add the product of coupling and probabilities, we get $-\frac{4}{3}g^2/(2\eta-1)$, very similar to the baryon case[5]. The negative sign indicates that the system is attractive because of the collective residual color coupling alone, even in the absence of volume pressure. This gives rise to a type of matter that is bound, but does not correspond to confined mesonic matter. We are interested in confined matter and will have to add a term to the energy to enforce this. In addition, we have

$$\vec{Q} = \sum_{i=1}^{2\eta} \vec{q}_i, \tag{1}$$

$$\vec{Q}^2 = \sum_{i=1}^{2\eta} \vec{q}_i^2 + 2\sum_{i \neq j} \vec{q}_i \cdot \vec{q}_j = 0. \tag{2}$$

This shows our model is an overall color singlet.

### B. TF differential equation

The total energy of a multi-quark meson can be calculated from Eq. (3). The first term gives kinetic energy, while the remaining are interaction terms.

$$
\begin{aligned}
E = &\sum_{i,I} \int^{r_{max}} d^3r \frac{\left(2\pi^2\hbar^3 N^I \hat{n}_i^I(r)\right)^{5/3}}{20\pi^2\hbar^3 m^I} + \\
&\frac{4}{3}g^2 \sum_I \frac{N^I\left(N^I-1\right)}{2} \int\int d^3r\, d^3r' \frac{\left(\sum_i P_{ii}\hat{n}_i^I(r)\hat{n}_i^I(r') - \frac{1}{2}\sum_{i<j} P_{ij}\hat{n}_i^I(r)\hat{n}_j^I(r')\right)}{|\vec{r}-\vec{r}'|} + \\
&\frac{4}{3}g^2 \sum_{I \neq J} \frac{N^I N^J}{2} \int\int d^3r\, d^3r' \frac{\left(\sum_i P_{ii}\hat{n}_i^I(r)\hat{n}_i^J(r') - \frac{1}{2}\sum_{i<j} P_{ij}\hat{n}_i^I(r)\hat{n}_j^J(r')\right)}{|\vec{r}-\vec{r}'|} + \\
&\frac{4}{3}g^2 \sum_I \frac{\bar{N}^I\left(\bar{N}^I-1\right)}{2} \int\int d^3r\, d^3r' \frac{\left(\sum_i \bar{\bar{P}}_{ii}\hat{n}_i^I(r)\hat{n}_i^I(r') - \frac{1}{2}\sum_{i<j} \bar{\bar{P}}_{ij}\hat{n}_i^I(r)\hat{n}_j^I(r')\right)}{|\vec{r}-\vec{r}'|} + \\
&\frac{4}{3}g^2 \sum_{I \neq J} \frac{\bar{N}^I \bar{N}^J}{2} \int\int d^3r\, d^3r' \frac{\left(\sum_i \bar{\bar{P}}_{ii}\hat{n}_i^I(r)\hat{n}_i^J(r') - \frac{1}{2}\sum_{i<j} \bar{\bar{P}}_{ij}\hat{n}_i^I(r)\hat{n}_j^J(r')\right)}{|\vec{r}-\vec{r}'|} - \\
&\frac{4}{3}g^2 \sum_{I,J} \bar{N}^I N^J \int\int d^3r\, d^3r' \frac{\left(\sum_i \bar{P}_{ii}\hat{n}_i^I(r)\hat{n}_i^J(r') - \frac{1}{2}\sum_{i<j} \bar{P}_{ij}\hat{n}_i^I(r)\hat{n}_j^J(r')\right)}{|\vec{r}-\vec{r}'|}.
\end{aligned}
\tag{3}
$$

In Eq. (3) the $I, J$ superscripts stand for flavor and the $i, j$ subscripts stand for color, and the $\hat{n}_i^I$ are number densities, which are normalized to one when integrated over space[4]. The number of quarks with flavor $I$ is designated as $N^I$. The number of antiquarks with flavor $I$ is designated as $\bar{N}^I$. From Table I, we can see that the probability of interaction type CCA is twice the CCR type. Also, interaction probability AAA is twice the AAR type. This simple finding amazingly removes four out of five terms from the interaction energy, leaving us with the last term from (3). Thus we conclude, for mesons on average, quarks only interact with antiquarks.

Assuming spherical symmetry, the TF spatial wave functions $f^I(r)$ and $\bar{f}^I(r)$ are related to the particle densities $n^I(r)$ and $\bar{n}^I(r)$ such that

$$f^I(r) = \frac{ra}{(2 \cdot 4\alpha_s/3)}(6\pi^2 n^I(r))^{2/3}, \tag{4}$$

and

$$\bar{f}^I(r) = \frac{ra}{(2 \cdot 4\alpha_s/3)}(6\pi^2 \bar{n}^I(r))^{2/3}, \tag{5}$$

where

$$a = \frac{\hbar}{m^1 c},  \tag{6}$$

gives the scale, where $m^1$ is the mass of lightest quark, and

$$\alpha_s = \frac{g^2}{\hbar c},  \tag{7}$$

is the strong coupling constant. Also, let's introduce the dimensionless parameter $x$ such that $r = Rx$ where

$$R = \frac{a}{(2 \cdot 4\alpha_s/3)} \left( \frac{3\pi\eta}{2} \right)^{2/3}.  \tag{8}$$

Variation of the densities $n^I(r)$ and $\bar{n}^I(r)$ over total energy, (i.e., Eq. (3)), and applying Eqs. (4), (5) and (8), we can obtain two second order differential equations, which are the TF quark equations for mesons. They are

$$\alpha^I \frac{d^2 f^I(x)}{dx^2} = -\frac{3\eta}{(2\eta-1)} \frac{1}{\sqrt{x}} \sum_{\bar{I}} \bar{f}^{\bar{I}}(x)^{3/2},  \tag{9}$$

$$\alpha^{\bar{I}} \frac{d^2 \bar{f}^{\bar{I}}(x)}{dx^2} = -\frac{3\eta}{(2\eta-1)} \frac{1}{\sqrt{x}} \sum_{I} f^I(x)^{3/2},  \tag{10}$$

where

$$\alpha^I = \frac{m^1}{m^I},  \tag{11}$$

is the ratio of the lightest quark to the $I^{th}$ quark. Equations (9) and (10) are the differential form of the TF quark equations in case of mesons. The interchangeability of these equations shows the TF equations are invariant with respect to particle and antiparticle. As was mentioned before, it also shows that quarks interact only with antiquarks in mesonic matter; quark/quark and antiquark/antiquark interactions sum to zero in the TF model. When there is an explicit particle/antiparticle symmetry, we assume $f = \bar{f}$ to reduce the TF differential equations into one:

$$\alpha^I \frac{d^2 f^I(x)}{dx^2} = -\frac{3\eta}{(2\eta-1)} \frac{1}{\sqrt{x}} \sum_{I} f^I(x)^{3/2},  \tag{12}$$

We use this equation to form system energies and equations for two of three different families of mesons in the sections below.

### 1. Charmonium family: case 1

Charmonium and all the multi-quark families of charmonium consists of charm and anti-charm only. Therefore, $I$ takes a single value and hence is dropped. $g_0$

refers to degeneracy and, for this application, can have the value of one or two.

$$\frac{d^2 f(x)}{dx^2} = -\frac{3\eta}{(2\eta-1)} \cdot g_0 \cdot \frac{1}{\sqrt{x}} f(x)^{3/2},  \tag{13}$$

where

$$g_0 \times N^I = \eta.  \tag{14}$$

We choose the normalization equation to be

$$\int_0^{x_{max}} dx \sqrt{x} f(x)^{3/2} = \frac{N^I}{3\eta}.  \tag{15}$$

Expressing the normalization equation in terms of boundary conditions, we get

$$(x\frac{df}{dx} - f)|_{x_{max}} = -\frac{\eta}{2\eta-1}.  \tag{16}$$

With these modified boundary conditions, we can derive the expression for kinetic $(T)$ and potential $(U)$ energies. For the kinetic energy we can start with[5]

$$T = 2 \cdot \sum_I \frac{12}{5\pi} \left( \frac{3\pi\eta}{2} \right)^{1/3} \frac{\frac{4}{3} g^2 \cdot \frac{4}{3}\alpha_s}{a} \times$$
$$\alpha_I \int_0^{x_I} dx \frac{(f^I(x))^{5/2}}{\sqrt{x}}.  \tag{17}$$

For a single flavor one has the simple result

$$T = 2 \cdot \frac{12}{5\pi} \left( \frac{3\pi\eta}{2} \right)^{1/3} \frac{\frac{4}{3} g^2 \cdot \frac{4}{3}\alpha_s}{a} \times$$
$$\left[ -\frac{5}{21} \frac{df(x)}{dx}|_{x_{max}} + \frac{4}{7} \sqrt{x_{max}} \left( f(x_{max}) \right)^{5/2} g_0 \right].  \tag{18}$$

Similarly, for potential energy we begin with

$$U = -\frac{9 \cdot \frac{4}{3} g^2}{(2\eta-1)} \times$$
$$\sum_{I,J} \int^{r_I} \int^{r_J} d^3 r \, d^3 r' \frac{n^I(r) n^J(r')}{|\vec{r} - \vec{r}'|}.  \tag{19}$$

Reducing this to TF wave functions, $f^I(x)$, we can get

$$U = -\frac{9 \cdot \frac{4}{3} g^2}{(2\eta-1)} \frac{\eta^2}{R} \times$$
$$\left( \sum_{I,J} \left[ \int_0^{x_I} dx \frac{(f^I(x))^{3/2}}{\sqrt{x}} \int_0^x dx' \sqrt{x'} \left( f^J(x') \right)^{3/2} \right. \right.$$
$$\left. \left. + \int_0^{x_I} dx (f^I(x))^{3/2} \sqrt{x} \int_x^{x_J} dx' \frac{(f^J(x'))^{3/2}}{\sqrt{x'}} \right] \right).  \tag{20}$$

Further simplification when there is a single flavor yields

$$U = \frac{4}{\pi} \left(\frac{3\pi\eta}{2}\right)^{1/3} \frac{\frac{4}{3}g^2 \cdot \frac{4}{3}\alpha_s}{a} \times$$
$$\left[\frac{4}{7}\frac{df(x)}{dx}|_{x_{max}} - \frac{4}{7}\sqrt{x_{max}}\left(f(x_{max})\right)^{5/2}g_0\right], \quad (21)$$

which like the expression for $T$ depends on the derivative and value of wave function at the boundary.

The volume energy ($E_v$) term gives the inward pressure which keeps quarks within a boundary. We assume that[7]

$$E_v = \frac{4}{3}\pi R^3 x_{max}^3 B, \quad (22)$$

where $B$ is the bag constant. Now that we have $T$, $U$ and $E_v$ terms, we can find the total energy of a desired multi-quark state. The total energy of such a state is simply given by

$$E = T + V + E_v + 2\eta \cdot m_c, \quad (23)$$

where $m_c$ is the mass of charm quark.

To assess the stability of multi-quark mesons we omit the trivial mass part and will examine the energy as a function of quark content.

### 2. Z-meson family: case 2

The constituents of Z-mesons are charm ($c$), anti-charm($\bar{c}$), light ($u$ or $d$) and anti-light quarks. We treat the mass of the up and down quarks as the same. This means the Z-meson and all multi-quark families of Z-meson have total quark mass equal to the antiquark mass. As before, we set $f = \bar{f}$ in the TF equations and obtain Eq. (12) with $I = 1, 2$. Let $f^1(x)$ be the wave function of the light quark and $f^2(x)$ be the wave function of the heavier quark. For $N_1$ quarks with degeneracy factor $g_1$, and $N_2$ quarks with degeneracy $g_2$, we have

$$g_1 N_1 + g_2 N_2 = \eta. \quad (24)$$

We assume a linear relation exists between $f^1(x)$ and $f^2(x)$ in the region $0 < x < x_2$, and that $f^2(x)$ vanishes for a dimensionless distance greater than $x_2$, i.e.,

$$f^1(x) = kf^2(x) \quad \text{for} \quad 0 \le x \le x_2,$$
$$f^2(x) = 0 \quad \text{for} \quad x_2 \le x \le x_1. \quad (25)$$

Eq. (12) can now be written as two set of equations:

$$\alpha^1 \frac{d^2 f^1(x)}{dx^2} = -\frac{3\eta}{(2\eta-1)}\frac{1}{\sqrt{x}}\left(g_1\left(f^1(x)\right)^{3/2}\right.$$
$$\left.+ g_2\left(f^2(x)\right)^{3/2}\right), \quad (26)$$

$$\alpha^2 \frac{d^2 f^2(x)}{dx^2} = -\frac{3\eta}{(2\eta-1)}\frac{1}{\sqrt{x}}\left(g_1\left(f^1(x)\right)^{3/2}\right.$$
$$\left.+ g_2\left(f^2(x)\right)^{3/2}\right). \quad (27)$$

To make Eqs. (26) and (27) consistent, we need

$$k = \frac{\alpha^2}{\alpha^1}, \quad (28)$$

which is just the inverse ratio of the given masses from (11). The similar step in the case of baryons gives a much more complicated consistency condition[5]. The normalization conditions are

$$\int_0^{x_2} x^{1/2}\left(f^2(x)\right)^{3/2}dx = \frac{N_2}{3\eta}, \quad (29)$$

and

$$\int_0^{x_1} x^{1/2}\left(f^1(x)\right)^{3/2}dx = \frac{N_1}{3\eta}. \quad (30)$$

In region $0 < x < x_2$,

$$\frac{d^2 f^1(x)}{dx^2} = Q_1 \frac{\left(f^1(x)\right)^{3/2}}{\sqrt{x}}, \quad (31)$$

where

$$Q_1 = -\frac{3\eta}{(2\eta-1)\alpha_1}\left(g_1 + \frac{g_2}{k^{3/2}}\right). \quad (32)$$

In region $x_2 < x < x_1$,

$$\frac{d^2 f^1(x)}{dx^2} = Q_2 \frac{\left(f^1(x)\right)^{3/2}}{\sqrt{x}}, \quad (33)$$

where

$$Q_2 = -\frac{3\eta}{(2\eta-1)\alpha_1}g_1. \quad (34)$$

With the equations above we can express normalization conditions in the form of boundary conditions:

$$\left(x\frac{df^2(x)}{dx} - f^2(x)\right)|_{x_2} = -\frac{N_2 Q_1 k^{3/2}}{3\eta}, \quad (35)$$

$$\left(x\frac{df^1(x)}{dx} - f^1(x)\right)|_{x_1} = -\frac{\eta}{(2\eta-1)\alpha_1}. \quad (36)$$

The energies are derived like in case 1. For $g_1$ flavors with $N_1$ particles and $g_2$ flavors with $N_2$ particles we have

$$T = 2 \cdot \frac{12}{5\pi}\left(\frac{3\pi\eta}{2}\right)^{1/3} \frac{\frac{4}{3}g^2 \cdot \frac{4}{3}\alpha_s}{a} \times$$
$$\left[g_1\alpha_1 \int_0^{x_1} \frac{\left(f^1(x)\right)^{5/2}}{\sqrt{x}}dx + \right.$$
$$\left. g_2\alpha_2 \int_0^{x_2} \frac{\left(f^2(x)\right)^{5/2}}{\sqrt{x}}dx\right]. \quad (37)$$

Using the wave function differential equations, the consistency condition for $k$ and boundary conditions allows one to relate the integrals to wave function values and derivatives on the surfaces:

$$
T = 2 \cdot \frac{12}{5\pi} \left( \frac{3\pi\eta}{2} \right)^{1/3} \frac{\frac{4}{3}g^2 \cdot \frac{4}{3}\alpha_s}{a} \times
$$

$$
\left[ -\frac{5}{21}\alpha_1 \frac{df^1(x)}{dx}\Big|_{x_1} + \frac{4}{7}\sqrt{x_1}\left(f^1(x_1)\right)^{5/2}g_1\alpha_1 \right.
$$

$$
\left. + \frac{4}{7}\sqrt{x_2}\left(f^2(x_2)\right)^{5/2}g_2\alpha_2 \right]. \tag{38}
$$

In the $(g_1, N_1), (g_2, N_2)$ case the expression for potential energy becomes

$$
U = -\frac{9 \cdot \frac{4}{3}g^2}{(2\eta-1)} \frac{\eta^2}{R} \left[ g_1^2 K_1 + g_2^2 K_2 + 2g_1 g_2 K_1 K_2 \right], \tag{39}
$$

where

$$
K_1 \equiv \int_0^{x_1} dx \frac{\left(f^1(x)\right)^{3/2}}{\sqrt{x}} \int_0^x dx' \sqrt{x'} \left(f^1(x')\right)^{3/2}
$$
$$
+ \int_0^{x_1} dx \left(f^1(x)\right)^{3/2} \sqrt{x} \int_x^{x_1} dx' \frac{\left(f^1(x')\right)^{3/2}}{\sqrt{x'}}, \tag{40}
$$

$$
K_2 \equiv \int_0^{x_2} dx \frac{\left(f^2(x)\right)^{3/2}}{\sqrt{x}} \int_0^x dx' \sqrt{x'} \left(f^2(x')\right)^{3/2}
$$
$$
+ \int_0^{x_2} dx \left(f^2(x)\right)^{3/2} \sqrt{x} \int_x^{x_2} dx' \frac{\left(f^2(x')\right)^{3/2}}{\sqrt{x'}}, \tag{41}
$$

$$
K_{12} \equiv \int_0^{x_1} dx \frac{\left(f^1(x)\right)^{3/2}}{\sqrt{x}} \int_0^x dx' \sqrt{x'} \left(f^2(x')\right)^{3/2}
$$
$$
+ \int_0^{x_1} dx \left(f^1(x)\right)^{3/2} \sqrt{x} \int_x^{x_2} dx' \frac{\left(f^2(x')\right)^{3/2}}{\sqrt{x'}}, \tag{42}
$$

and

$$
K_{21} \equiv \int_0^{x_2} dx \frac{\left(f^2(x)\right)^{3/2}}{\sqrt{x}} \int_0^x dx' \sqrt{x'} \left(f^1(x')\right)^{3/2}
$$
$$
+ \int_0^{x_2} dx \left(f^2(x)\right)^{3/2} \sqrt{x} \int_x^{x_1} dx' \frac{\left(f^1(x')\right)^{3/2}}{\sqrt{x'}}. \tag{43}
$$

We can see the $K_{12}$ integral is equivalent to the $K_{21}$ integral. After some further calculations, all these seemingly difficult integrals boil down to a simple equation, which is

$$
U = -\frac{4}{\pi} \left( \frac{3\pi\eta}{2} \right)^{1/3} \frac{\frac{4}{3}g^2 \cdot \frac{4}{3}\alpha_s}{a} \times
$$

$$
\left[ -\frac{4}{7}\alpha_1 \frac{df^1(x)}{dx}\Big|_{x_1} + \frac{4}{7}\sqrt{x_1}\left(f^1(x_1)\right)^{5/2}g_1\alpha_1 \right.
$$

$$
\left. + \frac{4}{7}\sqrt{x_2}\left(f^2(x_2)\right)^{5/2}g_2\alpha_2 \right]. \tag{44}
$$

One interesting thing we observed in both the kinetic and potential energy expressions is that there is no dependency on the derivative of wave function at the inner boundary, unlike the baryon case.

## 3. D-meson family: case 3

The D-meson and multi-quark families of D-mesons consist of charm and anti-light quarks. The heavier mass is always a quark and the lighter mass is always an antiquark in each member of the D-meson family. So, there is asymmetry in total mass of quarks and antiquarks. Also, because quarks interact only with antiquarks, the mathematics becomes slightly different than earlier cases. We have two different wave functions for the charm and antilight quark, $f$ and $\bar{f}$, respectively. We assume there is a universal wave function $\bar{f} = k_0 f$ in the region where the wave functions overlap. This means they are related linearly in the region where the wave function of charm quark is nonzero. Outside of this, only the wave function of the light antiquark exists.

For the region $0 < x < x_2$, we get the differential equation

$$
\frac{d^2 f(x)}{dx^2} = Q_0 \frac{\left(f(x)\right)^{3/2}}{\sqrt{x}}, \tag{45}
$$

where

$$
Q_0 = -\frac{3\eta \, g_0}{(2\eta-1) \cdot k_0}, \tag{46}
$$

and

$$
k_0 = \left(\frac{\alpha}{\bar{\alpha}}\right)^{2/5}, \tag{47}
$$

is the consistency condition. $\alpha$ is given by Eq. (11), and in this case $\bar{\alpha} = 1$.

In region $x_2 < x < x_1$ with two TF equations, one function is zero and the other becomes

$$
\frac{d^2 \bar{f}(x)}{dx^2} = 0,
$$
$$
\implies \bar{f}(x) = c \cdot x + d, \tag{48}
$$

where

$$
c = k_0 f'(x_2), \tag{49}
$$

and

$$d = k_0 \left( f(x_2) - f'(x_2) \cdot x_2 \right). \tag{50}$$

Proceeding as before we obtained expressions for kinetic and potential energies. The energies are

$$T = \frac{12}{5\pi} \left( \frac{3\pi\eta}{2} \right)^{1/3} \frac{\frac{4}{3}g^2 \cdot \frac{4}{3}\alpha_s}{a} \times$$
$$\left[ -\frac{10}{21}\alpha \frac{df(x)}{dx}|_{x_2} + \frac{8}{7}\sqrt{x_2}\left(f(x_2)\right)^{5/2} g_0 \alpha \right. \tag{51}$$
$$\left. + \int_{x_2}^{x_1} \frac{(cx+d)^{5/2}}{\sqrt{x}} dx \right],$$

$$U = -\frac{4}{\pi} \left( \frac{3\pi\eta}{2} \right)^{1/3} \frac{\frac{4}{3}g^2 \cdot \frac{4}{3}\alpha_s}{a} \times$$
$$\left[ -\frac{4}{7}\alpha \frac{df(x)}{dx}|_{x_2} + \frac{4}{7}\sqrt{x_2}\left(f(x_2)\right)^{5/2} g_0 \alpha \right. \tag{52}$$
$$\left. + \frac{\eta\, g_2}{2\eta - 1} \cdot \int_{x_2}^{x_1} \frac{(cx+d)^{3/2}}{\sqrt{x}} dx \right].$$

## III. METHOD AND REMARKS

The phenomenological parameters we used are the strong coupling constant $\alpha_s = 0.371$, bag constant $B^{1/4} = 74.5$ MeV, and light quark mass $m_1 = 300$ MeV. These came from a fit of the light baron spectrum[8]. In a future work we will instead attempt to fit the meson spectrum in this sector. We will examine wave functions and energies for the three cases. In nuclear physics, one examines the energy per nucleon in order to assess the stability of a given nucleus. Thus, the important figure of merit in these evaluations is the energy per quark, for if this increases as one adds more meson pairs, the family is unstable under decay to lower family members, whereas if it decreases, the family is stable. Note that we solved the differential equations using an iterative implementation of **NDSolve** in *Mathematica*.

## IV. RESULTS AND DISCUSSIONS

First, let us discuss the behavior of the density wave functions. The particle density function of charmonium, proportional to $(f(x)/x)^{3/2}$, drops with increase in distance and vanishes at a boundary as seen in Fig. 1.

The density function of Z-mesons has a long tail for the light quarks, while for charmed quarks the value is large and is concentrated near the origin, as seen in Fig. 2. This suggests an atomic-like structure with heavy charm, anti-charm quarks at the center while light quarks and antiquarks spread out like electrons. Fig. 3 is an enlargement of the density function of the light quark wave function for the Z-meson. It drops down abruptly until it reaches the boundary of the heavy quark wave function, then inflects and decreases. In the case of D-mesons, Fig. 4, the density function of light and heavy quarks are relatively closer. We increased the quark content and compared density functions of a family of multi-mesons in all three cases. We observed similar density functions for a given multi-meson family regardless of the quark content.

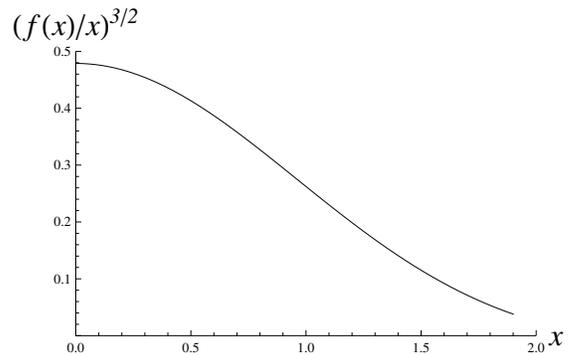

FIG. 1. Density function of charmonium.

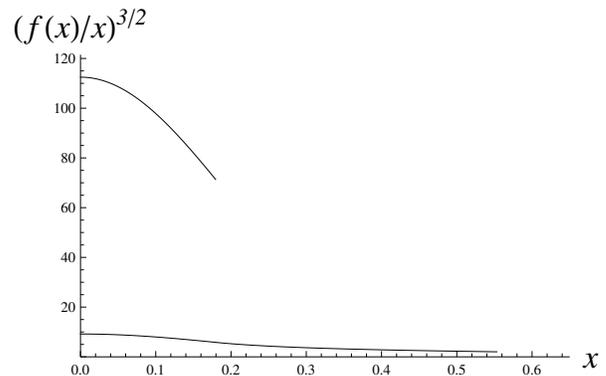

FIG. 2. Density function of Z-mesons.

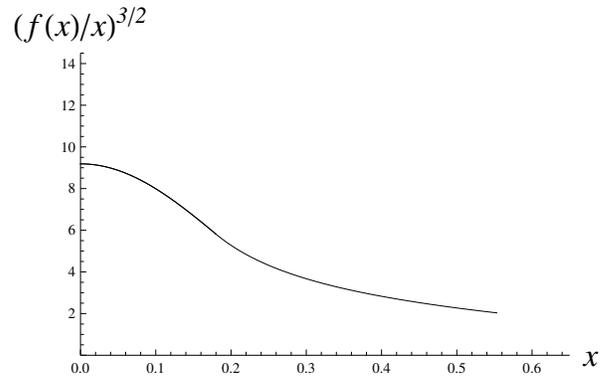

FIG. 3. Density function of light quarks for the Z-meson.

Fig. 5 includes 11 possibilities for the physical radius. The physical radius is plotted versus quark number and compared with a generic baryon with three degenerate

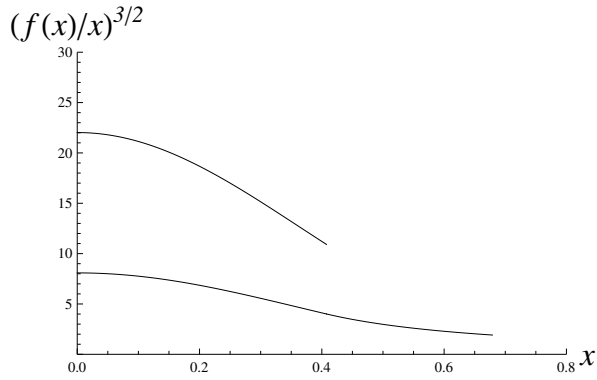

$(f(x)/x)^{3/2}$

FIG. 4. Density of wave functions for the D-meson.

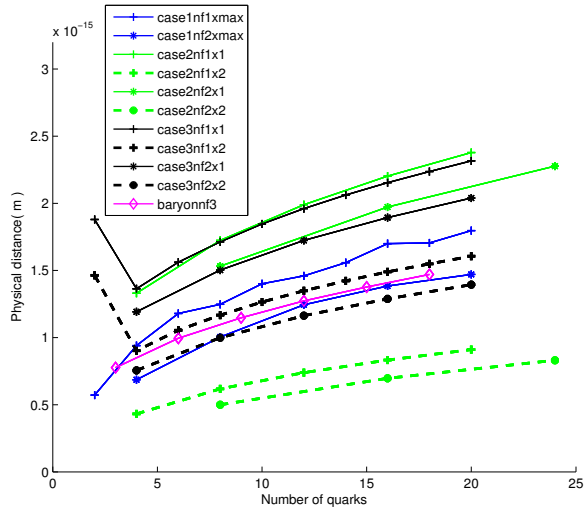

FIG. 5. Physical distance vs. quark content.

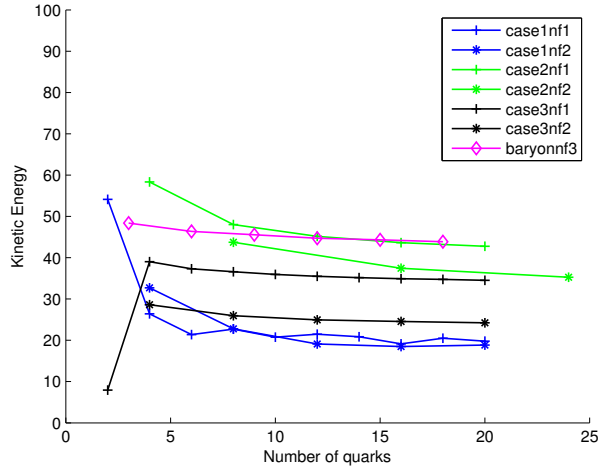

FIG. 6. Kinetic energy per quark vs. quark content.

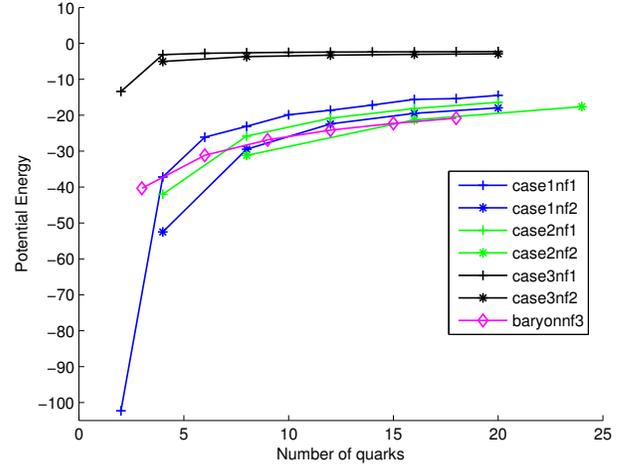

FIG. 7. Potential energy per quark vs. quark content.

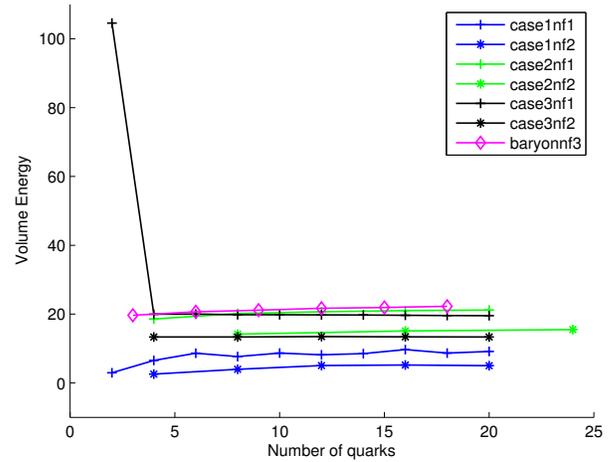

FIG. 8. Volume energy per quark vs. quark content.

light flavors. From Eq. (8) we can see that the radius is proportional to the product of $\eta^{2/3}$ and the dimensionless radius $x$. As the quark content increases, the dimensionless radius, $x$, becomes smaller, whereas $\eta^{2/3}$ increases. In most cases, the result is an increase in radius with increasing quark content. We also observe that the curve of the radius plot for each case tends to flatten out for larger numbers of quarks. **case1nf1xmax** refers to multi-quark families of charmonium with no degeneracy. In this case all the charm quarks have the same spin and hence cannot occupy the same state. **case1nf2xmax** is instead the plot of the charmonium family with a degeneracy of two. In this case, spin up and down is assigned to a pair of charm quarks, and thus they do not occupy the same state. The physical radius of **case1nf1xmax** being larger than **case1nf2xmax** reflects this fact. Note that the dotted lines refer to the inner boundary associated with the charmed quark in cases 2 and 3. For case

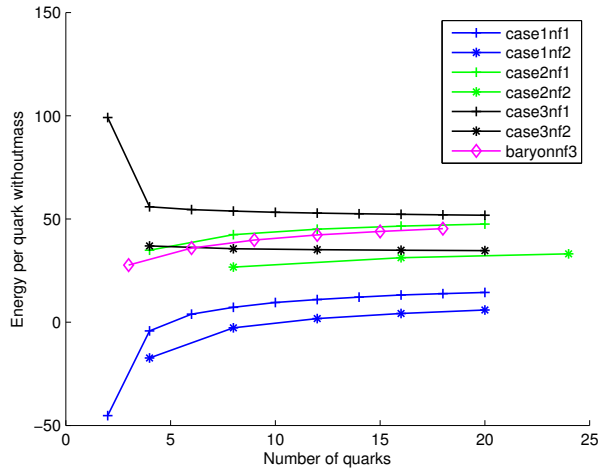

FIG. 9. Total energy per quark without mass term vs. quark number.

2 the difference between dotted and continuous lines is the largest. That means, like the Z-meson, all the higher quark family members have heavy charm-anticharm concentrated at the center while light and anti-light quarks are spread throughout. **case2nf1x2** refers to the radius plot of the inner boundary of the multi-quark family of Z-mesons with degeneracy of one while **case2nf2x2** refers to the same plot with degeneracy of two, and similarly for case 3. In all cases we see that the plot of physical radius is larger for degeneracy one compared to degeneracy two. The Z and D-meson family members are found to have equally large outer boundaries. For a given quark number, the outer radius of all types of mesons was found to be larger than the generic baryon except the degenerate case of the charmonium family. Note that these radii are determined by the size of the wave functions; the electromagnetic radii have not yet been evaluated.

There are 3 types of energies in this model: kinetic, potential and volume. The kinetic energy per quark (in MeV) depends strongly on the meson family, as seen in Fig. 6. We see that the energy per quark is relatively small and tends to decrease slowly, which seems to provide some justification for this nonrelativistic model. Fig. 7 shows the corresponding graph for the potential energy. We see that the D-meson family is remarkably different from the other mesons, and has the smallest potential energy per quark. Fig. 8 shows a huge jump downward in volume energy for the case 3 mesons from $n = 2$ to $n = 4$. The other energies are relatively flat.

Fig. 9 is our final result. It shows the total energy per quark without the mass term, i.e., the sum of kinetic, potential and volume energy, plotted against the quark content. The generic baryon rises slowly for increasing quark content, implying these are unstable; i.e., a higher quark content state can decay into lower members of the same family. The case 1 mesons rise quickly, and then continue the rise more slowly; these are also unstable. In contrast, the case 2 mesons do not rise so quickly but are rather similar to the baryon. However, for the case 3 family, the downward jump from $n = 2$ to $n = 4$ shows a tetraquark state which should be stable against decay into two of the lower family members. The other members continue a slow decrease, which also implies a family stability.

## V. CONCLUSIONS

We have initiated the study of multi-quark mesons using the TF quark model. After specifying the explicit interactions and summing on colors, we have found that quarks can only interact with antiquarks on average. We have also found that three cases of mesonic states may be constructed: charmonium family, Z-meson family and D-meson family. We have not yet included explicit spin interactions in our model, but we can take one level of degeneracy into account in our two-wave function construction. Using parameters fit in our baryonic study, we have observed interesting patterns of single-quark energies. Similar to our findings for baryons, the energy per quark is slowly rising for case 1 and 2 mesons, implying instability. However, we also found that a stable tetraquark state of the D-meson type is likely to exist. In addition, the other members show a possible family stability. These types of matter are rich and interesting and deserve more detailed study.

Further extensions of this model can be employed to examine states with central charge or mixed baryonic-mesonic states such as pentaquark families.

## VI. REFERENCES

[1] Z. Q. Liu et al., *Phys. Rev. Lett.* **110** (2013) 252002; S. K. Choi et al., *Phys. Rev. Lett.* **100** (2008) 142001.

[2] Ablikim, M. et al., *Phys. Rev. Lett.* **110** (2013) 252001; M. Ablikim et al., *Phys. Rev. Lett.* **115** (2015) 112003.

[3] R. Aaij et al., *Phys. Rev. Lett.* **118** (2017) 022003.

[4] W. Wilcox, *Nucl. Phys.* **A 826** (2009) 49.

[5] Q. Liu and W. Wilcox, *Ann. Phys.* **341** (2014) 164.

[6] P. Sikivie and N. Weiss, *Phys. Rev.* **D18** (1978) 3809.

[7] E. Farhi and R. Jaffe, *Phys. Rev.* **D30** (1984) 2379.

[8] The fit in Ref. 5 contained an error, but has been corrected; an *errata* will be submitted.